# On an algorithm that generates an interesting maximal set P(n) of the naturals for any n≥2

[1]Bidu Prakash Das and [2*]Soubhik Chakraborty

[1]Department of Computer Science and Engineering, BIT Mesra, Ranchi-835215, India
[2]Department of Applied Mathematics, BIT Mesra, Ranchi-835215, India

*God created the natural numbers. All the rest is the work of man.*

– Leopold Kronecker (1823-91)

**Abstract**

The paper considers the problem of finding the largest possible set P(n), a subset of the set N of the natural numbers, with the property that a number is in P(n) if and only if it is a sum of n distinct naturals all in P(n) or none in P(n). Here "largest" is in the set theoretic sense and n≥2. We call P(n) a maximal set obeying this property. For small n say 2 or 3, it is possible to develop P(n) intuitively but we strongly felt the necessity of an algorithm for any n≥2. Now P(n) shall invariably be a infinite set so we define another set Q(n) such that Q(n)=N-P(n), prove that Q(n) is finite and, since P(n) is automatically known if Q(n) is known, design an algorithm of worst case $O(1)$ complexity which generates Q(n).

The paper is organized as follows. Section 1 is the introduction where the algorithm is designed after a brief statement of the problem. Section 2 gives the code. Section 3 gives several examples with outputs of the code provided in the previous section. Finally section 4 is the conclusion and future work.

## 1. Introduction

The paper considers the problem of finding the largest possible set P(n), a subset of the set N of the natural numbers, with the property that a number is in P(n) if and only if it is a sum of n distinct integers all in P(n) or none in P(n). Here "largest" is in the set theoretic sense and n≥2. We call P(n) a maximal set obeying this property. For small n say 2 or 3, it is possible to develop P(n) intuitively [1] but we strongly felt the necessity of an algorithm for any n≥2. And this is precisely what is achieved here.

--------------------------------------------------------------------------------

*corresponding author's email: soubhikc@yahoo.co.in (S. Chakraborty)

Now P(n) shall invariably be a infinite set so we define the finite set Q(n) such that Q(n)=N-P(n) where the finiteness of Q(n) is also established in the paper. The finiteness of Q(n) implies that P(n) is automatically known if Q(n) is known, whence we design an algorithm of worst case O(1) complexity which generates Q(n).

Below we give the desired algorithm (the proof that Q(n) is finite is supplied in step 2):-

**The Algorithm**

**Step 1**

From the nature of the problem statement, for any given P(n), the first element in P(n) is the sum of first n natural numbers. Let X(1) represent the first element which is
1+ 2+ 3…. + n=n(n+1)/2. Now elements from 1 to ((n(n+1)/2)-1) i.e. till that natural number immediately preceding X(1) are not present in P(n) and are added to Q(n).

**Step 2**

Let z=X(1) + X(2) + X(3) +…..+ X(n-1) i.e. sum of first n-1 numbers belonging to P(n)
Let Y(1),Y(2)……………Y(z) represent the first z consecutive numbers all in P(n).
Now let S=Y(z) + 1 i.e. the number immediately following Y(z).
We must also have, S=Y(1) + z as the Y(i)'s are consecutive naturals.

As z is the sum of n-1 numbers all belonging to P(n) and Y(1) belongs to P(n) so S=Y(1)+z being the sum of n numbers all belonging to P(n) belongs to P(n).
Instead of using Y(1) if we use Y(2) till Y(z) then all numbers from Y(z) + 2 till Y(z) + z will belong to P(n). So now z more consecutive numbers belong to P(n). This z consecutive numbers can be used in a similar approach as above to include the next z consecutive numbers.

Proceeding like this all numbers greater than or equal to Y(1) belong to P(n).

**Theorem**: Q(n) is finite.

**Proof**: It is evident from above that we have obtained a finite number Y(1) beyond which all elements will be in P(n) so Q(n) is now a subset of elements from 1 to (Y(1) -1) and is thus a finite set.

**Step 3**

Our next task is to obtain the first z consecutive natural numbers all belonging to P(n). None of the numbers from 1 to X(1) -1 belong to P(n). These numbers taking n at a time can be used to generate all the natural numbers from X(1) ( X(1) is the smallest number generated taking first n natural numbers) till U where $U = (X(1) -1) + (X(1) -2) + (X(1) -3) \ldots (X(1) - n)$ i.e. the highest number possible taking the last n natural numbers of the series from 1 to $X(1) -1$.

As 1 does not belong to P(n), so $U + 1$ shall never belong to P(n) as it can only be generated using n numbers which belong to P(n) and one which does not. The next smallest number belonging to P(n) is $1 + 2 + 3 + \ldots\ldots\ldots(n-1) + (U+1)$
$=U1$ (say) as all n numbers do not belong to P(n).

All numbers between $U + 1$ and $U1 -1$ are not present in P(n) and are thus added to Q(n). The range of consecutive numbers belonging to P(n), possible from this new range of numbers none belonging to P(n) (from U+1 till U1 -1) are from U1 till U2
where  $U2 = 1 + 2 + 3 + 4 + \ldots\ldots n-1 + U1 -1$.

Now all numbers from X(1) till U belong to P(n). Using these numbers all numbers from U3 till U4 can be generated
where $U3 = X(1) + (X(1) + 1) + (X(1) + 2) + \ldots\ldots(X(1) + n-1)$
and    $U4 = U + (U -1) + (U-2) + \ldots\ldots\ldots(U-(n-1))$

If $U1 \leq U3$ and $U4 \geq U2$ then we have a bigger range of consecutive numbers.

Now if $(U4 - U1 + 1) \geq z$ our algorithm will hold true and all numbers greater than or equal to U1 will belong to P(n) else the new numbers generated are applied the same procedure as above till we get a sequence of consecutive numbers all belonging to P(n) having length greater than or equal to z.

The beauty of the problem in a way makes the problem trivial. A sequence of z consecutive numbers all belonging to P(n) is obtained in the first iteration itself. Also we have tested this for n>=3 and n<=10000 and have noticed that the difference between (U4 –U1) i.e. the length of consecutive numbers obtained in the first iteration and z increases with increasing values of n and thus for n>=10000 also it must hold true.

 For $n \geq 3$ the Algorithm amazingly reduces to a complexity **O(1).** This is because calculation of terms U1, U2, U3, U4 utilize only the value of n but the complexity in the steps involved are independent of this value. That is to say, irrespective of the value of n, only a constant number of steps are involved thus giving us a complexity of O(1).

For n=2 a consecutive sequence of 3 numbers (for n=2 z is found to be 3) all belonging to P(n) is found in the third iteration.

## 2. The CODE

The following is the code of our Algorithm written in C++ and run in GCC compiler

```
#include<iostream>
#include<conio.h>
using namespace std;
int main()
{
   unsigned long long int n,n1[100000],n2=0;
   cout<<"We are to find the range of numbers all belonging to the series as specified when given the value of n "<<endl<<endl;
   cout<<"Enter the value of n "<<endl<<endl;
   cin>>n;
   unsigned long long int c=0,i=1;
   for(i=1;i<=n;i++)
            c=c+i;
   cout<<"The first number in the series is ="<<c<<endl<<endl;
   //for(i=2;i<=n;i++)
   {
   //         cout<<" the "<<i<<"th number of the series is "<<c + i-1<<endl<<endl;
     //       getch();
   }
   cout<<"Now the following numbers are not in the series"<<endl<<endl;
   for(i=1;i<c;i++)
   {
            cout<<i<<" ";
            n1[n2++]=i;
   }
   cout<endl<<endl;
   cout<<"The range of consecutive numbers that can be formed using the above numbers all not belonging to the series, taking "<<n<<" at a time"<<endl<<endl;
   cout<<c<<"and ";
   unsigned long long int c1=0,c2=c,low=0,high=0;
   for(i=1;i<=n;i++)
   {
            c1=c-1 + c1;
            low=low + c2 + i -1;
            c--;
   }
   cout<<c1<<" "<<endl<<endl;
   cout<<"Using the above numbers and taking "<<n<<" at a time we can generate all numbers between and including ";
   unsigned long long int g66=0,g77=0;
   for(i=1;i<=n;i++)
   {
    g66=g66+ c2 + i-1;
    g77=g77 +c1 -i + 1;
   }
   cout<<g66<<" and "<<g77<<endl<<endl;
```

```
    for(i=1;i<=n;i++)
    high=high + c1 - i + 1;
    //cout<<" this is "<<low<<" "<<high<<endl;
    unsigned long long int g5=(n-1)*n;
    g5=g5/2;
    cout<<c1 + 1<<" will not be present in the series as it can be only formed by elements some of which are present in the series and some are not. "<<endl<<endl;
    cout<<"The next lowest number that can be formed is the sum of ";
    for(i=1;i<n;i++)
            cout<<i<<",";
    cout<<" and "<<c1 + 1<<"= "<<g5+ c1+1<<endl<<endl;
    cout<<"All numbers between "<<c1<<" and "<<g5+ c1+1<<" are not included in the series."<<endl<<endl;
    for(i=c1+1;i<g5 + c1 + 1;i++)
            n1[n2++]=i;
    unsigned long long int g=0,g1=c1+4;
    cout<<"Now the above numbers along with the numbers from 1 to "<<c2 -1 <<", all of which are not in the series can be used to generate all the numbers between the range "<<c1 + 1 +g5 <<" and ";
    for(i=1;i<n;i++)
            g=g+c2-i;
    g=g+3+c1;
    cout<<g<<endl<<endl;
    cout<<"So now we have "<<g-c1-g5 <<" consecutive numbers. "<<endl<<endl;
    cout<<"If we have "<<(n-1)*c2 + g5-n+1<<" consecutive numbers then all subsequent numbers will also be there."<<endl<<endl;
    cout<<"As we have "<<g<<" <" <<high<<" and "<<low<<" > "<<c1 + 1 +g5<<endl<<endl
    cout<<"The new range will be from "<<c1 + 1 +g5<<" to "<<high<<" giving us "<<high -c1 -g5<<" consecutive numbers"<<endl<<endl;
    cout<<"So now all numbers greater than "<<c1 + g5<<" are there in the given sequence."<<endl<<endl;
    cout<<"The numbers <= "<<c1 + g5<<" which are not present in the series gives us Q("<<"n "<<"):-"<<endl<<endl;

    for(i=0;i<n2;i++)
            cout<<n1[i]<<" ";
    cout<<endl<<endl;
}
```

## 3. Examples and computer output

**Ex.1:**
We are to find the range of numbers all belonging to the series as specified when given the value of n.

Enter the value of n  **4**

The first number in the series is =10.

Now the following numbers are not in the series:-

1 2 3 4 5 6 7 8 9

The range of numbers that can be formed using the above numbers all not belonging to the series, taking 4 at a time

10 and 30.

Using the above numbers and taking 4 at a time we can generate all numbers between and including 46 and 114.

31 will not be present in the series as it can be only formed by elements some of which are present in the series and some are not.

The next lowest number that can be formed is the sum of 1, 2, 3, and 31= 37.

All numbers between 30 and 37 are not included in the series.

Now the above numbers along with the numbers from 1 to 9, all of which are not in the series can be used to generate all the numbers between the range 37 and 57

So now we have 21 consecutive numbers.

If we have 33 consecutive numbers then all subsequent numbers will also be there.

As we have 57 <114 and 46 > 37

The new range will be from 37 to 114 giving us 78 consecutive numbers.

So now all numbers greater than 36 are there in the given sequence.

The numbers <= 36 which are not present in the series gives us Q(4):-

1 2 3 4 5 6 7 8 9 31 32 33 34 35 36

**Ex. 2**
We are to find the range of numbers all belonging to the series as specified when given the value of n

Enter the value of n **10**
The first number in the series is =55
Now the following numbers are not in the series:-

1 2 3 4 5 6 7 8 9 10 11 12 13 14 15 16 17 18 19 20 21 22 23 24 25 26 27 28 29 30 31 32 33 34 35 36 37 38 39 40 41 42 43 44 45 46 47 48 49 50 51 52 53 54

The range of numbers that can be formed using the above numbers all not belonging to the series, taking 10 at a time

55 and 495.

Using the above numbers and taking 10 at a time we can generate all numbers between and including 595 and 4905. 496 will not be present in the series as it can be only formed by elements some of which are present in the series and some are not.

The next lowest number that can be formed is the sum of 1, 2, 3, 4, 5, 6, 7, 8, 9, and 496= 541. All numbers between 495 and 541 are not included in the series.

Now the above numbers along with the numbers from 1 to 54, all of which are not in the series can be used to generate all the numbers between the range 541 and 948. So now we have 408 consecutive numbers. If we have 531 consecutive numbers then all subsequent numbers will also be there

As we have 948 <4905 and 595 > 541

The new range will be from 541 to 4905 giving us 4365 consecutive numbers.
So now all numbers greater than 540 are there in the given sequence.
The numbers less than or equal to 540 which are not present in the series gives us Q(10):-

1 2 3 4 5 6 7 8 9 10 11 12 13 14 15 16 17 18 19 20 21 22 23 24 25 26 27 28 29 30 31 32 33 34 35 36 37 38 39 40 41 42 43 44 45 46 47 48 49 50 51 52 53 54 496 497 498 499 500 501 502 503 504 505 506 507 508 509 510 511 512 513 514 515 516 517 518 519 520 521 522 523 524 525 526 527 528 529 530 531 532 533 534 535 536 537 538 539 540

**Ex. 3**
We are to find the range of numbers all belonging to the series as specified when given the value of n

Enter the value of n **40**

The first number in the series is =820.

Now the following numbers are not in the series:-

1 2 3 4 5 6 7 8 9 10 11 12 13 14 15 16 17 18 19 20 21 22 23 24 25 26 27 28 29 30 31 32 33 34 35 36 37 38 39 40 41 42 43 44 45 46 47 48 49 50 51 52 53 54 55 56 57 58 59 60 61 62 63 64 65 66 67 68 69 70 71 72 73 74 75 76 77 78 79 80 81 82 83 84 85 86 87 88 89 90 91 92 93 94 95 96 97 98 99 100 101 102 103 104 105 106 107 108 109 110 111

112 113 114 115 116 117 118 119 120 121 122 123 124 125 126 127 128 129 130 131
132 133 134 135 136 137 138 139 140 141 142 143 144 145 146 147 148 149 150 151
152 153 154 155 156 157 158 159 160 161 162 163 164 165 166 167 168 169 170 171
172 173 174 175 176 177 178 179 180 181 182 183 184 185 186 187 188 189 190 191
192 193 194 195 196 197 198 199 200 201 202 203 204 205 206 207 208 209 210 211
212 213 214 215 216 217 218 219 220 221 222 223 224 225 226 227 228 229 230 231
232 233 234 235 236 237 238 239 240 241 242 243 244 245 246 247 248 249 250 251
252 253 254 255 256 257 258 259 260 261 262 263 264 265 266 267 268 269 270 271
272 273 274 275 276 277 278 279 280 281 282 283 284 285 286 287 288 289 290 291
292 293 294 295 296 297 298 299 300 301 302 303 304 305 306 307 308 309 310 311
312 313 314 315 316 317 318 319 320 321 322 323 324 325 326 327 328 329 330 331
332 333 334 335 336 337 338 339 340 341 342 343 344 345 346 347 348 349 350 351
352 353 354 355 356 357 358 359 360 361 362 363 364 365 366 367 368 369 370 371
372 373 374 375 376 377 378 379 380 381 382 383 384 385 386 387 388 389 390 391
392 393 394 395 396 397 398 399 400 401 402 403 404 405 406 407 408 409 410 411
412 413 414 415 416 417 418 419 420 421 422 423 424 425 426 427 428 429 430 431
432 433 434 435 436 437 438 439 440 441 442 443 444 445 446 447 448 449 450 451
452 453 454 455 456 457 458 459 460 461 462 463 464 465 466 467 468 469 470 471
472 473 474 475 476 477 478 479 480 481 482 483 484 485 486 487 488 489 490 491
492 493 494 495 496 497 498 499 500 501 502 503 504 505 506 507 508 509 510 511
512 513 514 515 516 517 518 519 520 521 522 523 524 525 526 527 528 529 530 531
532 533 534 535 536 537 538 539 540 541 542 543 544 545 546 547 548 549 550 551
552 553 554 555 556 557 558 559 560 561 562 563 564 565 566 567 568 569 570 571
572 573 574 575 576 577 578 579 580 581 582 583 584 585 586 587 588 589 590 591
592 593 594 595 596 597 598 599 600 601 602 603 604 605 606 607 608 609 610 611
612 613 614 615 616 617 618 619 620 621 622 623 624 625 626 627 628 629 630 631
632 633 634 635 636 637 638 639 640 641 642 643 644 645 646 647 648 649 650 651
652 653 654 655 656 657 658 659 660 661 662 663 664 665 666 667 668 669 670 671
672 673 674 675 676 677 678 679 680 681 682 683 684 685 686 687 688 689 690 691
692 693 694 695 696 697 698 699 700 701 702 703 704 705 706 707 708 709 710 711
712 713 714 715 716 717 718 719 720 721 722 723 724 725 726 727 728 729 730 731
732 733 734 735 736 737 738 739 740 741 742 743 744 745 746 747 748 749 750 751
752 753 754 755 756 757 758 759 760 761 762 763 764 765 766 767 768 769 770 771
772 773 774 775 776 777 778 779 780 781 782 783 784 785 786 787 788 789 790 791
792 793 794 795 796 797 798 799 800 801 802 803 804 805 806 807 808 809 810 811
812 813 814 815 816 817 818 819.

The range of numbers that can be formed using the above numbers all not belonging to the series, taking 40 at a time are

820 and 31980

Using the above numbers and taking 40 at a time we can generate all numbers between and including 33580 and 1278420. 31981 will not be present in the series as it can be only formed by elements some of which are present in the series and some are not.

The next lowest number that can be formed is the sum of
1,2,3,4,5,6,7,8,9,10,11,12,13,14,15,16,17,18,19,20,21,22,23,24,25,26,27,28,29,30,31,32,
33,34,35,36,37,38,39, and 31981= 32761. All numbers between 31980 and 32761 are not
included in the series.

Now the above numbers along with the numbers from 1 to 819, all of which are not in the
series can be used to generate all the numbers between the range 32761 and 63183
So now we have 30423 consecutive numbers. If we have 32721 consecutive numbers
then all subsequent numbers will also be there

As we have 63183 <1278420 and 33580 > 32761

The new range will be from 32761 to 1278420 giving us 1245660 consecutive numbers
So now all numbers greater than 32760 are there in the given sequence. The numbers less
than or equal to 32760 which are not present in the series gives us Q(40):-

1 2 3 4 5 6 7 8 9 10 11 12 13 14 15 16 17 18 19 20 21 22 23 24 25 26 27 28 29 30 31 32
33 34 35 36 37 38 39 40 41 42 43 44 45 46 47 48 49 50 51 52 53 54 55 56 57 58 59 60
61 62 63 64 65 66 67 68 69 70 71 72 73 74 75 76 77 78 79 80 81 82 83 84 85 86 87 88
89 90 91 92 93 94 95 96 97 98 99 100 101 102 103 104 105 106 107 108 109 110 111
112 113 114 115 116 117 118 119 120 121 122 123 124 125 126 127 128 129 130 131
132 133 134 135 136 137 138 139 140 141 142 143 144 145 146 147 148 149 150 151
152 153 154 155 156 157 158 159 160 161 162 163 164 165 166 167 168 169 170 171
172 173 174 175 176 177 178 179 180 181 182 183 184 185 186 187 188 189 190 191
192 193 194 195 196 197 198 199 200 201 202 203 204 205 206 207 208 209 210 211
212 213 214 215 216 217 218 219 220 221 222 223 224 225 226 227 228 229 230 231
232 233 234 235 236 237 238 239 240 241 242 243 244 245 246 247 248 249 250 251
252 253 254 255 256 257 258 259 260 261 262 263 264 265 266 267 268 269 270 271
272 273 274 275 276 277 278 279 280 281 282 283 284 285 286 287 288 289 290 291
292 293 294 295 296 297 298 299 300 301 302 303 304 305 306 307 308 309 310 311
312 313 314 315 316 317 318 319 320 321 322 323 324 325 326 327 328 329 330 331
332 333 334 335 336 337 338 339 340 341 342 343 344 345 346 347 348 349 350 351
352 353 354 355 356 357 358 359 360 361 362 363 364 365 366 367 368 369 370 371
372 373 374 375 376 377 378 379 380 381 382 383 384 385 386 387 388 389 390 391
392 393 394 395 396 397 398 399 400 401 402 403 404 405 406 407 408 409 410 411
412 413 414 415 416 417 418 419 420 421 422 423 424 425 426 427 428 429 430 431
432 433 434 435 436 437 438 439 440 441 442 443 444 445 446 447 448 449 450 451
452 453 454 455 456 457 458 459 460 461 462 463 464 465 466 467 468 469 470 471
472 473 474 475 476 477 478 479 480 481 482 483 484 485 486 487 488 489 490 491
492 493 494 495 496 497 498 499 500 501 502 503 504 505 506 507 508 509 510 511
512 513 514 515 516 517 518 519 520 521 522 523 524 525 526 527 528 529 530 531
532 533 534 535 536 537 538 539 540 541 542 543 544 545 546 547 548 549 550 551
552 553 554 555 556 557 558 559 560 561 562 563 564 565 566 567 568 569 570 571
572 573 574 575 576 577 578 579 580 581 582 583 584 585 586 587 588 589 590 591
592 593 594 595 596 597 598 599 600 601 602 603 604 605 606 607 608 609 610 611
612 613 614 615 616 617 618 619 620 621 622 623 624 625 626 627 628 629 630 631

632 633 634 635 636 637 638 639 640 641 642 643 644 645 646 647 648 649 650 651
652 653 654 655 656 657 658 659 660 661 662 663 664 665 666 667 668 669 670 671
672 673 674 675 676 677 678 679 680 681 682 683 684 685 686 687 688 689 690 691
692 693 694 695 696 697 698 699 700 701 702 703 704 705 706 707 708 709 710 711
712 713 714 715 716 717 718 719 720 721 722 723 724 725 726 727 728 729 730 731
732 733 734 735 736 737 738 739 740 741 742 743 744 745 746 747 748 749 750 751
752 753 754 755 756 757 758 759 760 761 762 763 764 765 766 767 768 769 770 771
772 773 774 775 776 777 778 779 780 781 782 783 784 785 786 787 788 789 790 791
792 793 794 795 796 797 798 799 800 801 802 803 804 805 806 807 808 809 810 811
812 813 814 815 816 817 818 819 31981 31982 31983 31984 31985 31986 31987 31988
31989 31990 31991 31992 31993 31994 31995 31996 31997 31998 31999 32000 32001
32002 32003 32004 32005 32006 32007 32008 32009 32010 32011 32012 32013 32014
32015 32016 32017 32018 32019 32020 32021 32022 32023 32024 32025 32026 32027
32028 32029 32030 32031 32032 32033 32034 32035 32036 32037 32038 32039 32040
32041 32042 32043 32044 32045 32046 32047 32048 32049 32050 32051 32052 32053
32054 32055 32056 32057 32058 32059 32060 32061 32062 32063 32064 32065 32066
32067 32068 32069 32070 32071 32072 32073 32074 32075 32076 32077 32078 32079
32080 32081 32082 32083 32084 32085 32086 32087 32088 32089 32090 32091 32092
32093 32094 32095 32096 32097 32098 32099 32100 32101 32102 32103 32104 32105
32106 32107 32108 32109 32110 32111 32112 32113 32114 32115 32116 32117 32118
32119 32120 32121 32122 32123 32124 32125 32126 32127 32128 32129 32130 32131
32132 32133 32134 32135 32136 32137 32138 32139 32140 32141 32142 32143 32144
32145 32146 32147 32148 32149 32150 32151 32152 32153 32154 32155 32156 32157
32158 32159 32160 32161 32162 32163 32164 32165 32166 32167 32168 32169 32170
32171 32172 32173 32174 32175 32176 32177 32178 32179 32180 32181 32182 32183
32184 32185 32186 32187 32188 32189 32190 32191 32192 32193 32194 32195 32196
32197 32198 32199 32200 32201 32202 32203 32204 32205 32206 32207 32208 32209
32210 32211 32212 32213 32214 32215 32216 32217 32218 32219 32220 32221 32222
32223 32224 32225 32226 32227 32228 32229 32230 32231 32232 32233 32234 32235
32236 32237 32238 32239 32240 32241 32242 32243 32244 32245 32246 32247 32248
32249 32250 32251 32252 32253 32254 32255 32256 32257 32258 32259 32260 32261
32262 32263 32264 32265 32266 32267 32268 32269 32270 32271 32272 32273 32274
32275 32276 32277 32278 32279 32280 32281 32282 32283 32284 32285 32286 32287
32288 32289 32290 32291 32292 32293 32294 32295 32296 32297 32298 32299 32300
32301 32302 32303 32304 32305 32306 32307 32308 32309 32310 32311 32312 32313
32314 32315 32316 32317 32318 32319 32320 32321 32322 32323 32324 32325 32326
32327 32328 32329 32330 32331 32332 32333 32334 32335 32336 32337 32338 32339
32340 32341 32342 32343 32344 32345 32346 32347 32348 32349 32350 32351 32352
32353 32354 32355 32356 32357 32358 32359 32360 32361 32362 32363 32364 32365
32366 32367 32368 32369 32370 32371 32372 32373 32374 32375 32376 32377 32378
32379 32380 32381 32382 32383 32384 32385 32386 32387 32388 32389 32390 32391
32392 32393 32394 32395 32396 32397 32398 32399 32400 32401 32402 32403 32404
32405 32406 32407 32408 32409 32410 32411 32412 32413 32414 32415 32416 32417
32418 32419 32420 32421 32422 32423 32424 32425 32426 32427 32428 32429 32430
32431 32432 32433 32434 32435 32436 32437 32438 32439 32440 32441 32442 32443
32444 32445 32446 32447 32448 32449 32450 32451 32452 32453 32454 32455 32456

32457 32458 32459 32460 32461 32462 32463 32464 32465 32466 32467 32468 32469
32470 32471 32472 32473 32474 32475 32476 32477 32478 32479 32480 32481 32482
32483 32484 32485 32486 32487 32488 32489 32490 32491 32492 32493 32494 32495
32496 32497 32498 32499 32500 32501 32502 32503 32504 32505 32506 32507 32508
32509 32510 32511 32512 32513 32514 32515 32516 32517 32518 32519 32520 32521
32522 32523 32524 32525 32526 32527 32528 32529 32530 32531 32532 32533 32534
32535 32536 32537 32538 32539 32540 32541 32542 32543 32544 32545 32546 32547
32548 32549 32550 32551 32552 32553 32554 32555 32556 32557 32558 32559 32560
32561 32562 32563 32564 32565 32566 32567 32568 32569 32570 32571 32572 32573
32574 32575 32576 32577 32578 32579 32580 32581 32582 32583 32584 32585 32586
32587 32588 32589 32590 32591 32592 32593 32594 32595 32596 32597 32598 32599
32600 32601 32602 32603 32604 32605 32606 32607 32608 32609 32610 32611 32612
32613 32614 32615 32616 32617 32618 32619 32620 32621 32622 32623 32624 32625
32626 32627 32628 32629 32630 32631 32632 32633 32634 32635 32636 32637 32638
32639 32640 32641 32642 32643 32644 32645 32646 32647 32648 32649 32650 32651
32652 32653 32654 32655 32656 32657 32658 32659 32660 32661 32662 32663 32664
32665 32666 32667 32668 32669 32670 32671 32672 32673 32674 32675 32676 32677
32678 32679 32680 32681 32682 32683 32684 32685 32686 32687 32688 32689 32690
32691 32692 32693 32694 32695 32696 32697 32698 32699 32700 32701 32702 32703
32704 32705 32706 32707 32708 32709 32710 32711 32712 32713 32714 32715 32716
32717 32718 32719 32720 32721 32722 32723 32724 32725 32726 32727 32728 32729
32730 32731 32732 32733 32734 32735 32736 32737 32738 32739 32740 32741 32742
32743 32744 32745 32746 32747 32748 32749 32750 32751 32752 32753 32754 32755
32756 32757 32758 32759 32760

**4.0 Conclusions and future work**

We claim to have designed an algorithm which generates the maximal set of naturals $P(n)$ with the property that a number is in $P(n)$ if it is the sum of n distinct naturals either all in $P(n)$ or none in $P(n)$ for any $n \geq 2$. Future work involves linking this problem with Ramsey theory[2] which has several applications in social sciences.

**Acknowledgement**

The first author wishes to acknowledge the encouragement he received from **Dr. Shamir Khuller** and an anonymous referee who read his popular science note [3] and encouraged him to extend the same to a research problem.

**References**

1. S. Chakraborty, A Curious Set of Numbers, Resonance, Nov. 2003, p. 94-95

2. R. L. Graham, B. L. Rothschild, J. H. Spencer, Ramsey Theory, 2$^{nd}$ Edition, John Wiley and Sons, 1990